\documentclass{article}

\usepackage{microtype}
\usepackage{graphicx}
\usepackage{subfigure}
\usepackage{booktabs} 

\usepackage{hyperref}

\usepackage[accepted]{icml2025}

\usepackage{amsmath}
\usepackage{amssymb}
\usepackage{mathtools}
\usepackage{amsthm}

\usepackage{amssymb}
\usepackage{color}
\usepackage{colortbl}
\usepackage{multirow}
\usepackage{makecell}
\usepackage{bbding}
\usepackage{wrapfig}
\usepackage{amsfonts}
\usepackage{pifont} 
\newcommand{\cmark}{\ding{51}}
\newcommand{\xmark}{\ding{55}}%
\usepackage{diagbox} 
\usepackage{xcolor}
\newcommand{\modified}[1]{\textcolor{black}{#1}} 

\usepackage[capitalize,noabbrev]{cleveref}

\theoremstyle{plain}

\theoremstyle{definition}

\theoremstyle{remark}

\usepackage[textsize=tiny]{todonotes}

\icmltitlerunning{SongGen: A Single Stage Auto-regressive Transformer for Text-to-Song Generation}

\begin{document}

\twocolumn[
\icmltitle{SongGen: A Single Stage Auto-regressive Transformer \\for Text-to-Song Generation}




\begin{icmlauthorlist}
\icmlauthor{Zihan Liu}{buaa,shlab}
\icmlauthor{Shuangrui Ding}{cuhk}
\icmlauthor{Zhixiong Zhang}{hit}
\icmlauthor{Xiaoyi Dong}{shlab,cuhk}
\icmlauthor{Pan Zhang}{shlab}
\icmlauthor{Yuhang Zang}{shlab}
\icmlauthor{Yuhang Cao}{shlab}
\icmlauthor{Dahua Lin}{cuhk,shlab,cpii}
\icmlauthor{Jiaqi Wang}{shlab}
\end{icmlauthorlist}

\icmlaffiliation{buaa}{Beihang University, Beijing, China}
\icmlaffiliation{shlab}{Shanghai AI Laboratory, Shanghai, China}
\icmlaffiliation{cuhk}{The Chinese University of Hong Kong, Hong Kong, China}
\icmlaffiliation{hit}{Harbin Institute of Technology, Harbin, China}
\icmlaffiliation{cpii}{CPII under InnoHK, Hong Kong, China}

\icmlcorrespondingauthor{Jiaqi Wang}{wangjiaqi@pjlab.org.cn}

\icmlkeywords{}

\vskip 0.3in
]



\printAffiliationsAndNotice{}  

\begin{abstract}

Text-to-song generation, the task of creating vocals and accompaniment from textual inputs, poses significant challenges due to domain complexity and data scarcity. 
\modified{Existing approaches often employ multi-stage generation procedures, leading to cumbersome training and inference pipelines, as well as suboptimal overall generation quality due to error accumulation across stages.}
In this paper, we propose \textbf{SongGen}, a fully open-source, single-stage auto-regressive transformer designed for controllable song generation. The proposed model facilitates fine-grained control over diverse musical attributes, including lyrics and textual descriptions of instrumentation, genre, mood, and timbre, while also offering an optional three-second reference clip for voice cloning. Within a unified auto-regressive framework, SongGen supports two output modes: \textbf{mixed mode}, which generates a mixture of vocals and accompaniment directly, and \textbf{dual-track mode}, which synthesizes them separately for greater flexibility in downstream applications. We explore diverse token pattern strategies for each mode, leading to notable improvements and valuable insights. Furthermore, we design an automated data preprocessing pipeline with effective quality control. To foster community engagement and future research, we will release our model weights, training code, annotated data, and preprocessing pipeline.
The code is available at \url{https://github.com/LiuZH-19/SongGen}.

\end{abstract}

\section{Introduction}
Songs, blending vocals with instrumental accompaniment, are a cornerstone of musical expression. Unlike purely instrumental music, songs uniquely capture human emotions through emotive lyrics and diverse melodies. However, creating a song is a complex, multi-stage process involving composition, instrumental arrangement, vocal performance, and more. This process requires substantial time and expertise, making it challenging for most individuals.
With the rise of AI Generated Content (AIGC), creative fields have been revolutionized, extending from text and image generation~\cite{rombach2022high, zhang2023adding,achiam2023gpt} to sophisticated artistic domains like music~\cite{huang2018music, dhariwal2020jukebox,ji2020comprehensive}. Building on these advancements, text-to-song generative models aim to transform natural language descriptions into full-song audio, making music creation more accessible and efficient.

\begin{figure}[tb!]
	\centering
	\includegraphics[width=0.96\columnwidth]{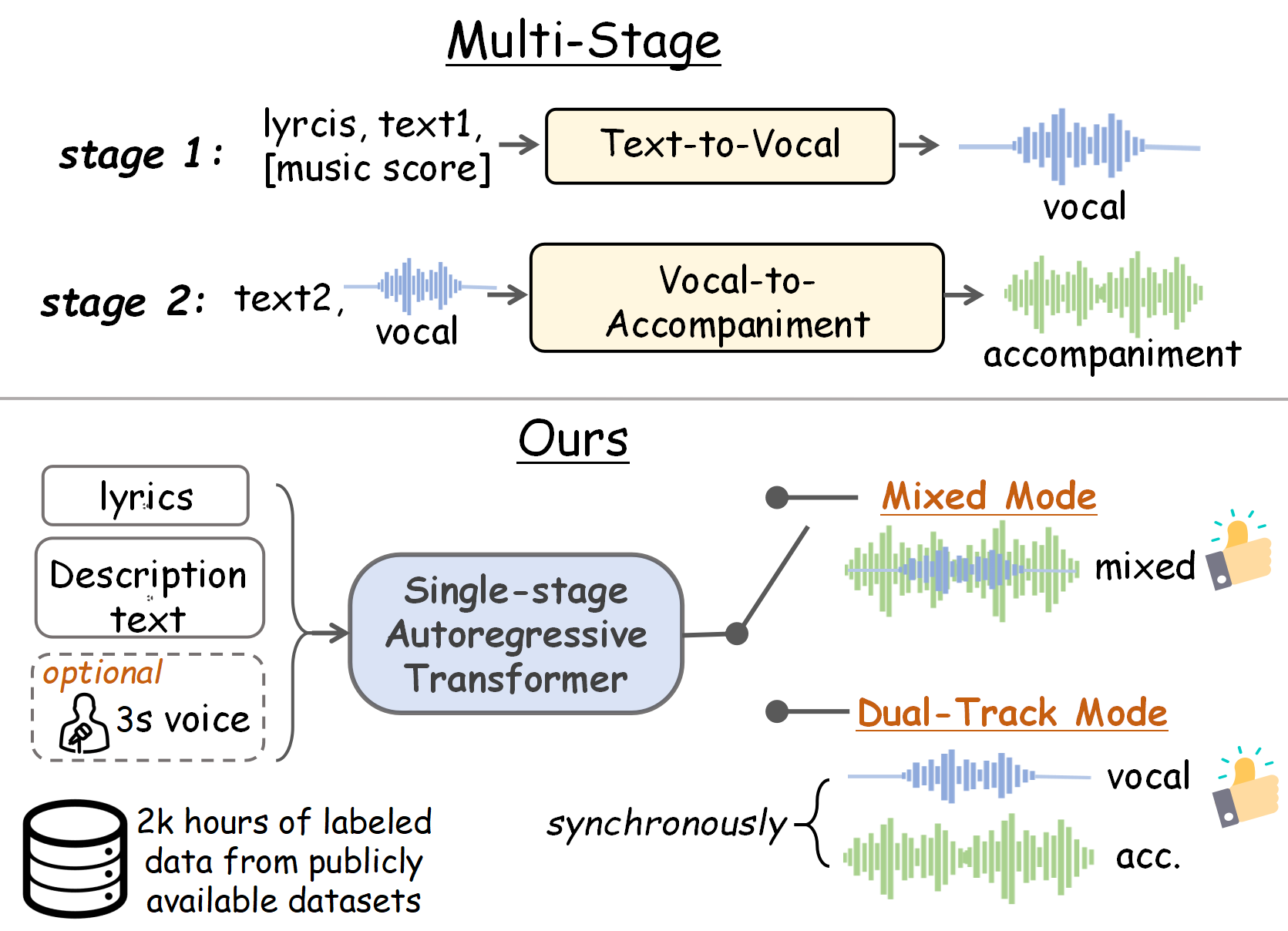}
    \vspace{-10pt}
	\caption{
    Multi-stage approaches generate vocals and accompaniment sequentially, leading to cumbersome pipelines and cumulative errors. SongGen simplifies this with a single-stage auto-regressive transformer that jointly models both tracks, supporting mixed and dual-track modes with better quality and efficiency.
    }
	\label{fig:motivation}
    \vspace{-10pt}
\end{figure}

Song generation presents greater complexity than speech or instrumental music generation~\cite{lyth2024parler, chen2024valle2valle2,liu2024audioldm, copet2024musicgen}. 
Unlike speech, singing spans a broader pitch range, incorporates a wider variety of expressive vocal techniques, and follows more dynamic rhythmic patterns. Moreover, song generation requires precise alignment between vocal and instrumental components to ensure musicality, harmonic consistency, and structural coherence—making it a uniquely challenging and underexplored problem.
The scarcity of open-source data further limits research in this area.

Recent text-to-song approaches~\cite{hong2024Melodist, li2024melodyLM} decompose songs into separate vocal and accompaniment tracks and adopt multi-stage generation pipelines. As illustrated in Figure~\ref{fig:motivation}, these models first generate the vocal track from lyrics, then produce the accompaniment using natural language prompts alongside the generated vocals.
However, such pipeline-based approach often fail to capture global optimality due to error accumulation across stages. This limitation is especially problematic for song generation. For instance, in genres like rap, vocal rhythm is tightly coupled with the instrumental beat. Generating vocals first without considering the underlying rhythm may result in rhythm misalignment. Conversely, in expressive genres such as ballads, where vocals typically guide the emotional flow, generating accompaniment first may constrain vocal expressiveness, resulting in rigid or disconnected performances. In both cases, pipeline approaches struggle to capture the intricate interplay between vocals and accompaniment. 
In addition, multi-stage generation results in cumbersome training and inference pipelines. 
Given these limitations in generation quality and efficiency, an important question arises: Is it possible for a single-stage model to achieve effective text-to-song generation?

In this paper, we introduce SongGen, a fully open-source, single-stage text-to-song generation model based on an auto-regressive transformer architecture. SongGen transforms lyrics and descriptive text into songs with harmonized vocals and accompaniment, allowing fine-grained control over instruments, genre, mood, timbre, and other musical elements. With a three-second reference vocal clip, it also supports zero-shot voice cloning. These user-defined controls are incorporated through modal-specific encoders, learnable projectors, and cross-attention mechanisms.  
SongGen offers two flexible generation modes: \textbf{mixed mode}, which blends vocals and accompaniment into a single output, and \textbf{dual-track mode}, which synthesizes them separately to facilitate professional post-production editing.

However, due to the sophisticated relationship between vocals and accompaniment in a song, jointly predicting them with natural expressiveness is a non-trivial task. To this end, we perform extensive explorations into output token patterns, yielding valuable insights. Specifically, (1) in \textbf{mixed mode}, while the model generates high-quality accompaniment, it struggles with natural-sounding vocals. Accompaniment, with higher energy and stable spectral distribution, tends to converge faster during training, whereas vocals, with higher semantic density and a lower signal-to-noise ratio due to overlap, present greater modeling challenges. This learning bias makes it difficult to generate vocals with clear lyrics, a problem typically addressed by decoupling and multi-stage methods. To mitigate this issue, we introduce an auxiliary vocal token prediction target, enhancing the model’s focus on vocal features and significantly improving vocal clarity in mixed-token outputs. (2) In \textbf{dual-track mode}, vocals and accompaniment are treated as distinct yet interconnected sequences, generated in sync by a single transformer decoder. We explore various track combination patterns to maintain precise frame-level alignment. Experimental results indicate that the optimal pattern yields well-coordinated vocals and accompaniment, achieving quality on par with mixed-mode generation.

Moreover, the text-to-song generation community has long been constrained by data scarcity. To the best of our knowledge, no publicly available dataset currently includes paired audio, lyrics, and captions. To bridge this gap, we develop an automated pipeline for data cleaning, processing, and quality filtering, resulting in a high-quality dataset of 540K song clips spanning over 2,000 hours of audio. 

To evaluate the effectiveness of the proposed SongGen framework, we conduct extensive experiments on the MusicCaps~\cite{agostinelli2023musiclm} test set.
The results demonstrate that SongGen outperforms the multi-stage baseline and generates songs with excellent musicality and vocal-instrument harmony, achieving performance that is competitive with the ground truth. 
Surprisingly, the generated songs feature expressive vocal techniques, such as vibrato, enhancing naturalness and authenticity.
Our contributions can be summarized as follows:
\vspace{-5pt}
\begin{itemize}
\setlength{\itemsep}{0pt}
 \item We introduce SongGen, a single-stage auto-regressive transformer for text-to-song generation, offering versatile control via lyrics, descriptive text, and an optional reference voice.
 \item SongGen supports both mixed and dual-track mode to accommodate diverse requirements. Our experiments provide valuable insights for optimizing both modes.
 \item By releasing the model weights, code, annotated data, and preprocessing pipeline, we aim to establish a simple yet effective baseline for future song generation research.
\end{itemize}

\section{Related Work} 
\begin{figure*}[tb!]
	\centering
	\includegraphics[width=1.9\columnwidth]{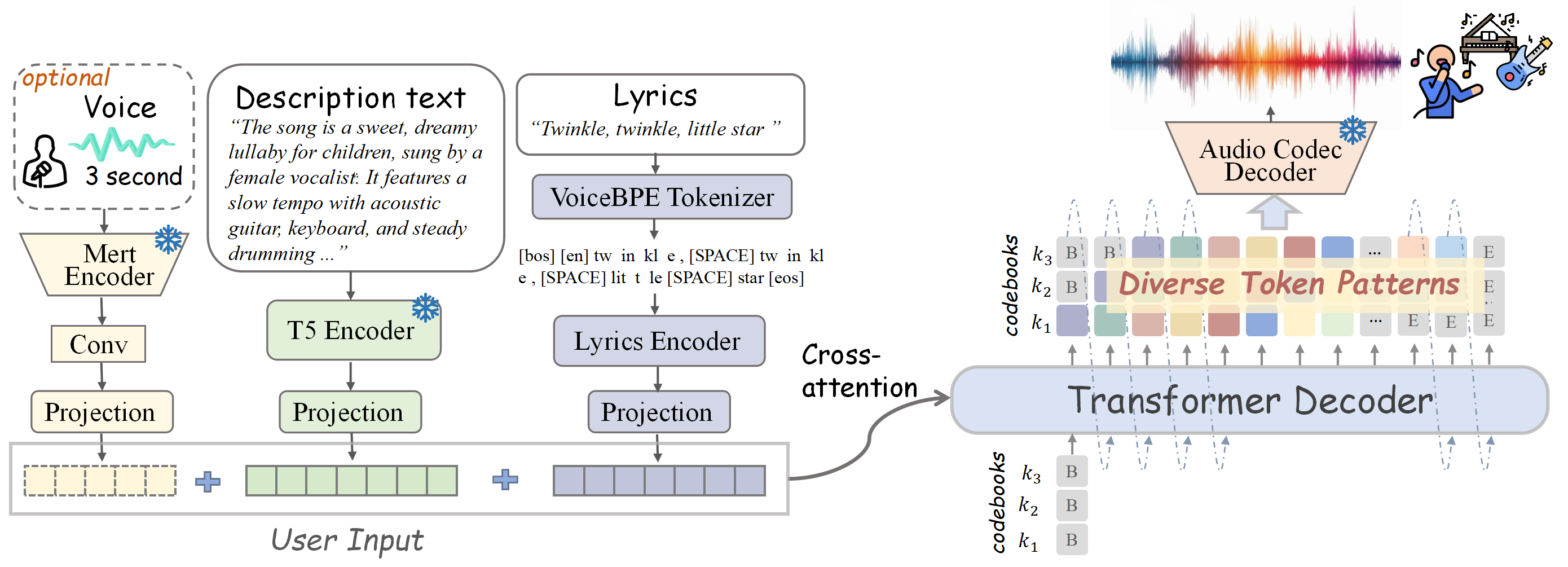}
    \vspace{-10pt}
	\caption{Overview of SongGen: An auto-regressive transformer decoder generates audio tokens with diverse patterns, incorporating user-defined controls via cross-attention. The final song is synthesized from these tokens through the audio codec decoder.
    }
	\label{fig:framework}
    \vspace{-10pt}
\end{figure*}

\subsection{Text-to-Music Generation}
In recent years, significant progress has been made in text-to-music generation models, which use descriptive text as a condition for controllable music generation.
Several works \cite{agostinelli2023musiclm, copet2024musicgen} employ transformer-based language models (LMs) \citep{vaswani2017attention} to model sequences of discrete tokens derived from audio codecs \citep{defossez2022encodec, zeghidour2021soundstream,yang2023hifi, kumar2024DAC}.
Diffusion models~\citep{sohl2015deep, ho2020denoising, kingma2021variational}, another competitive class of generative models, have also attained impressive results in music generation\cite{forsgren2022riffusion, chen2024musicldm, evans2024stable,schneider2023mo, huang2023noise2music, liu2024audioldm}.
\modified{While most models generate a mixture of stems, MusicGen-Stem~\cite{rouard2025musicgenstem} proposes a multi-stem generative model with three separate tracks (bass, drums and other), suggesting that disentangling components can facilitate generation and source editing.}
However, although all the models discussed above excel at generating high-quality instrumental music, they face significant challenges in producing realistic vocals.

\subsection{Song Generation}
Recently, a few studies have begun exploring song generation, a task that involves vocal composition, instrumental arrangement, and harmonious generation.
One of the pioneering efforts is Jukebox \cite{dhariwal2020jukebox}, which employs a multi-scale VQ-VAE to compress audio into discrete codes and models them using a cascade of transformer models.
However, Jukebox offers limited style control, relying solely on genre tags and artist names, and suffers from long inference times.
Recently, models like Melodist \cite{hong2024Melodist} and MelodyLM \cite{li2024melodyLM} have adopted multi-stage approaches to address the challenges of text-to-song generation. Melodist integrates singing voice synthesis with vocal-to-accompaniment (V2A) techniques, while MelodyLM improves upon Melodist by overcoming its reliance on music scores through a three-stage process: text-to-MIDI, text-to-vocal, and V2A. However, both approaches result in cumbersome training and inference procedures, and their corpus is limited to Mandarin pop songs, lacking diversity.
Another model, SongCreator \cite{lei2024songcreator}, utilizes a dual-sequence language model to capture the relationship between vocals and accompaniment. However, it lacks text-based control and produces vocals with limited clarity. 
Freestyle \cite{ning2024freestyler} focuses on generating rapping vocals from lyrics and accompaniment inputs but is constrained to a single musical style, with rap typically featuring simpler melodies.
\modified{The concurrent work Yue~\cite{yuan2025yue} achieves impressive results by scaling up both data and model size, adopting a track-decoupled next-token prediction and a two-stage causal LM framework. In contrast, we explore more diverse and efficient token pattern designs within a single-stage transformer.}
Although industry tools like Suno\footnote{\url{https://suno.com/}} and Udio\footnote{\url{https://www.udio.com/}} have recently emerged for song generation, neither has disclosed their methodologies or expanded into broader controllable generation tasks. SeedMusic \cite{bai2024seedmusic} leverages both auto-regressive language modeling and diffusion approaches to support song generation. However, SeedMusic is not open-source and relies on a large proprietary dataset, making a fair comparison with our fully open model unfeasible.

\section{Methodology}

\subsection{Overview}

\begin{figure*}[tb!]
	\centering
	\includegraphics[width=1.99\columnwidth]{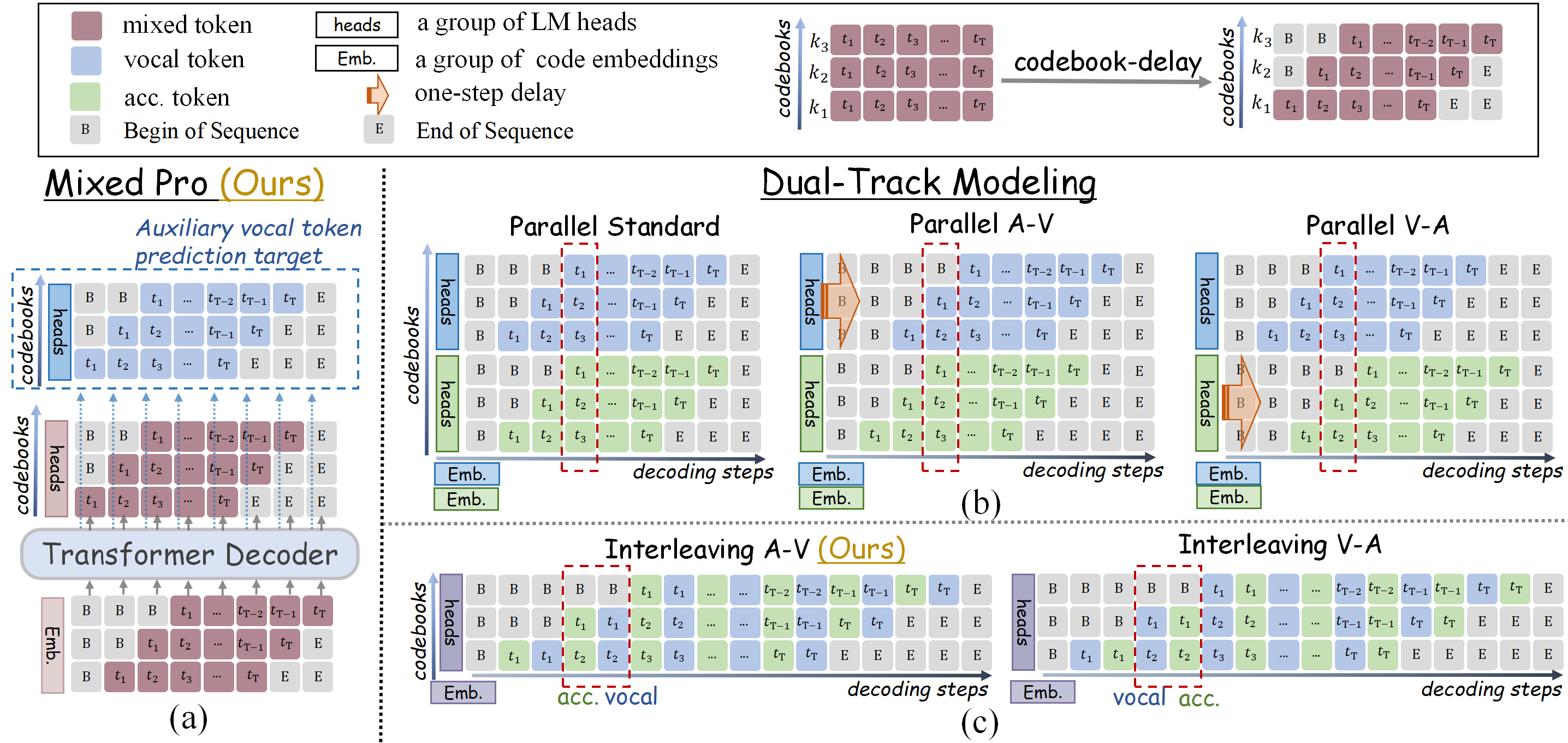}
    \vspace{-10pt}
	\caption{Illustration of token patterns for different generation modes. The codebook-delay pattern (from MusicGen) is applied to every audio token. (a) \textbf{Mixed Pro}: Directly decoding mixed tokens, with an auxiliary vocal token prediction target to enhance vocal learning.  \textbf{Dual-track mode:} (b) Parallel: Vocal and accompaniment tokens are concatenated along the codebook dimension, with three track order variants. (c) Interleaving: Tokens from both tracks are interleaved along the temporal dimension, with two track order variants.}
	\label{fig:pattern}
    \vspace{-10pt}
\end{figure*}

The objective of this paper is to guide the generation of a song using a text description, lyrics, and an optional reference voice.
As illustrated in Figure \ref{fig:framework}, SongGen is composed of an auto-regressive transformer decoder with an off-the-shelf neural audio codec. The transformer decoder predicts a sequence of audio tokens, allowing control through user inputs via cross-attention. The final song is synthesized from these tokens using the codec decoder.
In the subsequent section, we will elaborate on the details of SongGen. Section~\ref{sec:mode} will introduce the two generation modes supported by our unified framework: mixed mode and dual-track mode. Section~\ref{sec:cond_model} will discuss the lyric, voice, and text conditions. Section~\ref{sec:pipeline} will outline our data processing pipeline and quality filtering metrics. Section~\ref{sec:train} will present our training scheme for progressively enhancing model performance.

\subsection{Auto-regressive Codec Language Modeling}
\label{sec:mode}
\subsubsection{Audio tokenization}
The effectiveness of the audio tokenizer is critical to the success of transformer-based song generation.
Our framework is compatible with mainstream Codec designs.
In experiments, we employ X-Codec \cite{ye2024xcodec}, an audio codec based on Residual Vector Quantizer (RVQ)~\cite{zeghidour2021soundstream}, to produce discrete audio tokens.
It utilizes $N_{q}=8$ codebooks, each with a codebook size of $K=1024$. Given an audio signal $X \in \mathbb{R}^{d \cdot f_s}$, where $d$ is the audio duration and $f_s = 16$ kHz is the sampling rate, X-Codec encodes and quantizes $X$ into a sequence of token vectors $\mathbf{S} = [\mathbf{s}^1, \mathbf{s}^2, \dots, \mathbf{s}^T] \in \mathbb{R}^{N_q \times T}$, where $T= d\cdot f_r$ and $f_r=50$ HZ is the frame rate. Each vector $\mathbf{s}^t = [s^{1,t}, s^{2,t}, \dots, s^{N_q, t}]$ consists of $N_q$ codes, with $s^{k,t}$ taking integer values from 0 to $K-1$ for $k \in [1, N_q]$. 
We apply the codebook-delay pattern~\cite{copet2024musicgen} to handle the multiple code sequences within a single transformer decoder architecture. Figure \ref{fig:pattern} at the top-right corner illustrates this process for the case of $N_q = 3$, where a one-step delay is maintained between adjacent sequences from different codebooks. After applying the delay pattern, the resulting code sequences are denoted as $\hat{S}  \in \mathbb{R}^{N_q \times T'}$.

\subsubsection{Mixed Mode}

In mixed mode generation, we directly use the mixed audio tokens $\hat{\mathbf{S}}_{\text{mixed}}$, which are encoded by X-Codec from mixed audio (i.e. raw audio), as the output target.
For each step, the vector of audio tokens 
from $N_q$ codebooks are embedded using a group of $N_q$ learnable embedding matrices, and then summed up to form the decoder input. Additionally, a sinusoidal positional embedding is added at each step.
The last hidden state of decoder is passed to a group of $N_q$ linear heads, with each head predicting the logits corresponding to its respective codebook.

During training, we employ the teacher-forcing scheme. 
Since each quantizer in the RVQ encodes the quantization error from the previous quantizer, earlier codebooks are more critical. Therefore, we compute a weighted sum of the losses from different codebooks, assigning higher importance to the losses from earlier codebooks:
\begin{equation}
\begin{aligned} 
\label{eq:mix_loss}
\mathcal{L}_{\text{mixed}} &= \sum_{k=1}^{N_q} w_k \cdot \mathcal{L}^k_{\text{mixed}}, 
\end{aligned}    
\end{equation}
where $k$ denotes the codebook index, and $w_k$ represents the weight, satisfying $w_k \leq w_j$ for $ k < j $ and $ \sum_{k=1}^{N_q} w_k=1$. $\mathcal{L}^k_{\text{mixed}}$ is the cross-entropy loss for the $k$-th codebook.

However, this basic approach, referred to as ``Mixed", presents challenges in producing coherent and clear vocals. 
In mixed audio, vocals suffer from a low signal-to-noise ratio because of overlap with the accompaniment. While the accompaniment typically exhibits higher energy and a more stable spectral distribution, the vocals tend to be sparser, more irregular, and prone to greater instantaneous frequency fluctuations. For example, vocals often feature rapid pitch changes to perform various singing techniques. Moreover, vocals carry more semantic meaning from the lyrics. When mixed audio is used as the training target, the model tends to prioritize the more predictable accompaniment, often neglecting the vocal features. Nevertheless, human perception is sensitive to the naturalness and clarity of vocals, making these aspects critically important in song generation.

Building on this, we propose a method called ``Mixed Pro" that emphasizes vocal learning by introducing an auxiliary vocal token prediction target. 
As depicted in Figure \ref{fig:pattern} (a), we incorporate a dedicated group of linear heads to predict logits for vocal tokens. These tokens, encoded by X-Codec from the vocal track, are aligned frame-by-frame with the mixed tokens. 
The overall loss function is formulated as:
\begin{equation}
\begin{aligned} 
\mathcal{L}_{\text{mixed-pro}} &=  \mathcal{L}_{\text{mixed}} + \lambda \mathcal{L}_{\text{vocal}},
\end{aligned}    
\end{equation}
where $\lambda$  controls the contribution of the vocal loss to the total loss.
It is important to note that these newly introduced vocal heads are used only during training to compute the auxiliary loss and do not affect inference.

\subsubsection{Dual-track Mode}
In dual-track generation mode, the two key components of a song—the vocal and the accompaniment—are separated, and SongGen synchronously generates both tracks within this unified framework.
Considering the importance of harmony between vocals and accompaniment, we introduce two combination patterns, namely Parallel and Interleaving, to ensure frame-level alignment across the two tracks.

\textbf{Parallel:}
Inspired by the stereo channel modeling of MusicGen~\cite{copet2024musicgen}, which simultaneously outputs audio tokens for two channels, we design a parallel pattern. As shown in Figure \ref{fig:pattern} (b), the accompaniment and vocal audio tokens are concatenated along the codebook dimension, with each step containing $N_q$ vocal tokens and $N_q$ accompaniment tokens. On the temporal dimension, we introduce three variants. In the ``Standard" variant, the audio tokens for both tracks are strictly aligned frame by frame. The ``Parallel (A-V)" variant delays the vocal tokens by one step relative to the accompaniment tokens. Thus, the vocal token prediction at each frame considers both the previous vocal token and the accompaniment token at the current frame. 
Conversely, in the ``Parallel (V-A)" variant, the accompaniment tokens are delayed by one step relative to the vocal tokens. 
Two groups of code embeddings are used to separately embed the audio tokens for the two tracks. All embeddings are then averaged to form a combined input. Two groups of linear heads are employed to predict the audio tokens for each track. The training loss is defined as:
\begin{equation}
\label{eq:dual_loss}
\begin{aligned}
\mathcal{L}_{\text{parallel}} &= \frac{1}{2} (\mathcal{L}_{\text{vocal}} +  \mathcal{L}_{\text{acc}}),
\end{aligned}    
\end{equation}
where  $\mathcal{L}_{\text{vocal}} $ and $ \mathcal{L}_{\text{acc}} $ represent the individual losses for the vocal and accompaniment tracks, respectively. The calculation method is the same as in Equation \ref{eq:mix_loss}.

\textbf{Interleaving:}
In this pattern, the audio tokens of the two tracks are interleaved along the temporal dimension, as illustrated in Figure \ref{fig:pattern} (c). There are two variants: ``Interleaving (A-V)", where the accompaniment tokens precede the vocal tokens at each frame; and ``Interleaving (V-A)", where the vocal tokens precede the accompaniment tokens. 
In the "Interleaving (A-V)" variant, each vocal token prediction at a given frame considers both the previous vocal token and the accompaniment token from the same frame, with the reverse for the "Interleaving (V-A)" variant.
In this pattern, only a single group of code embeddings and one group of heads are used. The training loss is calculated in the same way as in Equation \ref{eq:dual_loss}.

Although the interleaving pattern requires longer sequence lengths than the parallel pattern, it provides a more effective approach to modeling the relationship between vocals and accompaniment. In the lower layers of the transformer, the interleaving pattern facilitates learning the interactions between the vocal and accompaniment tracks, while the higher layers focus on refining the distinct characteristics of each track. The attention visualizations in Figure \ref{fig:attan} provide additional evidence for this. In contrast, the parallel pattern is unable to decouple the vocal and accompaniment information before reaching the heads.

\subsection{Model Conditioning}
\label{sec:cond_model}
\textbf{Lyrics Conditioning.}
To address the challenge of data scarcity, we apply a 6681-token voice Byte-Pair Encoding (VoiceBPE) tokenizer \cite{casanova2024xtts} to convert the lyrics $C_{\text{lyrics}}$ into a sequence of phoneme-like tokens. Word-level tokenizers, like the T5\cite{raffel2020T5} tokenizer, lead to sparse training samples for each token embedding. In contrast, VoiceBPE not only enhances the model's ability to generalize to unseen words but also adapts more effectively to the variations in phoneme duration and pitch range inherent in sung vocals. Subsequently, the lyrics embedding $E_{\text{lyrics}} \in \mathbb{R}^{T_l \times F_l}$ is obtained by passing the lyric tokens through a small transformer-based encoder (i.e., Lyrics Encoder) to extract critical pronunciation-related information. Here, $T_l$ denotes the length of the lyric tokens, and $F_l$ represents the dimensionality of the embedding.

\textbf{Voice Conditioning. }
As demonstrated by the Marble~\cite{yuan2023marble} benchmark, MERT~\cite{li2024mert}, a music representation model, consistently achieves state-of-the-art performance in vocal technique detection and singer identification tasks. Consequently, we employ a frozen MERT encoder to generate robust voice feature embeddings, enabling control over vocal timbre and singing techniques.
Specifically, we randomly select 3-second clips from vocal segments to serve as the voice condition input, denoted as $C_{\text{voice}}$. 
The outputs from MERT's 24 hidden layers and 1 output layer are aggregated via a 1D convolutional layer, yielding the voice embedding $E_{\text{voice}} \in \mathbb{R}^{T_v \times F_v}$, where $T_v$ denotes the temporal length and $F_v$ represents the feature dimensionality of the embedding.

\textbf{Text Conditioning. }
Our text descriptions cover a wide range of musical attributes, including but not limited to the instruments used, musical emotion, tempo, genre, and the singer’s gender, offering more depth than simple tags or short phrases.
Given a description $C_{\text{text}}$  matching the song, we apply a frozen FLAN-T5~\cite{chung2022flanT5} encoder to obtain the text embedding, denoted as $E_{\text{text}} \in \mathbb{R}^{T_t \times F_t}$.

The above three condition embeddings—$E_{\text{lyrics}}$, $ E_{\text{voice}} $, and $ E_{\text{text}} $—are each passed through their respective projection layers to obtain transformed embeddings,  $\hat{E}_{\text{lyrics}} $, $ \hat{E}_{\text{voice}} $, and $\hat{E}_{\text{text}} $. These embeddings are then concatenated along the temporal dimension:
\begin{align}
 E_{\text{cond}} = \hat{E}_{\text{voice}} \oplus  \hat{E}_{\text{text}}  \oplus \hat{E}_{\text{lyrics}} \in \mathbb{R}^{(T_v+T_l+T_t) \times D},   
\end{align}
where $D$ denotes the dimension of the decoder hidden states.
This concatenated embedding \( E_{\text{cond}} \) is used to control song generation via cross attention.

\subsection{Automated Data Preprocessing Pipeline}
\label{sec:pipeline}
To the best of our knowledge, there is currently no publicly available dataset for text-to-song generation that includes paired audio, lyrics, and captions. To address this gap, we develop an automated data annotation pipeline that incorporates several filtering strategies to ensure high-quality data. 
\textbf{(1) Data Source:} We collect 8,000 hours of audio from Million Song Dataset (MSD) \cite{Bertin2011msd}, Free Music Archive (FMA) \cite{fma_challenge} and MTG-Jamendo Dataset \cite{bogdanov2019mtg}.
\textbf{(2) Source Separation:} We utilize Demucs \citep{rouard2022demucs} to separate vocals and accompaniment from the original audio. 
\textbf{(3) Segmentation:} We employ a voice activity detection (VAD) tool~\cite{gao2023funasr} to detect voiced segments in the separate vocal tracks. 
Vocal, accompaniment, and mixed tracks are then sliced according to the VAD results, with an average clip duration of 15 seconds. 
Additionally, the energy of each clip is calculated as the sum of the squared amplitude over time, providing a measure of loudness. Clips with low energy in either the accompaniment or vocals are discarded.
\textbf{(4) Lyric Recognition:} Lyric recognition accuracy is crucial for song generation, but it is challenging. Existing Automatic Speech Recognition (ASR) models, trained on speech data, struggle with the complexity and variability of sung vocals. 
Errors arise from two main factors: ASR limitations (misrecognitions and hallucinations); and inherently unclear vocal data, such as noise or genre-specific characteristics like those in rock music.
To tackle this issue, we apply two ASR models, Whisper-large-v2 and Whisper-larger-v3 ~\citep{radford2022whisper}, to automatically transcribe the vocals and generate two lyric transcriptions. We compute the edit distance between them to assess quality, excluding clips with an edit distance greater than 20\%, and retaining only those with relatively clearer vocals and higher recognition confidence.
\textbf{(5) Captioning:} 
We use LP-MusicCaps-MSD~\cite{doh2023lp} for MSD captions. For song clips without captions, we generate pseudo-captions using a music captioning model~\citep{doh2023lp}. The accuracy of the captions is evaluated by CLAP Score, which measures the alignment between audio and text with the official CLAP\cite{laionclap2023} model. Samples with low CLAP scores are discarded, and any available original tags are added as a supplement.
After preprocessing, the training dataset contains about 540K English-voiced clips, totaling around 2K hours of audio.

\subsection{Training Scheme}
\label{sec:train}
\textbf{Mixed Mode Training. }
Our mixed mode training consists of three key steps, aimed at progressively boost model performance. \textit{Step 1: Modality Alignment.} We train the entire model using total paired data to align the modalities between the various conditioning inputs and the audio output. 
\textit{Step 2 : Voice-Free Support.} 
To enable the model to function without a reference voice, we apply a 50\% random drop to the reference voice input. To maintain the model's original capabilities, we freeze all modules related to user inputs and fine-tune only the transformer decoder. Once the decoder adapts, we unfreeze the entire model and fine-tune all parameters to optimize performance.
\textit{Step 3: High-Quality Fine-tuning.} The final stage refines the model using a carefully selected subset of data filtered by these quality metrics: $\text{edit\_distance} \leq 5\%$, $\text{CLAP}_{src} \geq 25\%$, $\text{energy} > 1000$. This yields 100K high-quality pairs for fine-tuning, enabling the model to enhance the quality of audio by learning from cleaner, more relevant data.

\textbf{Dual-track Mode Training. }
Our experiments revealed that training the dual-track mode from scratch is challenging. To address this, we initialize the dual-track model with the pre-trained mixed mode model after Step 1. \textit{Step 1.5: Dual-Track Mode Adaptation.} After initialization, we freeze user input modules and fine-tune only the transformer decoder to adapt it to the new token pattern. Once the adaptation is complete, we unfreeze all model weights and proceed to fine-tune the entire model. The subsequent training steps mirror those of Steps 2 and 3 in the mixed mode.

\textbf{Curriculum Learning for Codebook Loss Weight Adjustment. }
We propose a curriculum learning strategy to adjust the weights of codebook losses during training. Initially, the first three codebooks have weights of 0.25, while the rest are set to 0.05. 
 This encourages the model to focus on the most important components first. As training progresses, the weights are gradually balanced, enabling the model to capture finer audio details step by step.

\section{Experiments}
\subsection{Experimental setup}

\begin{table*}[t]
\caption{Automatic evaluation of Text-to-Song generation. * denotes that we finetune Parlet-tts using our training data. The top two results, excluding the ground truth, are marked in \textbf{bold} and \underline{underlined}, respectively. } 
\label{table:t2s_auto}
\begin{center}
\resizebox{1.9\columnwidth}{!}{
\begin{tabular}{l|lcccccccccc}
\toprule

\multicolumn{2}{l}{\multirow{2}{*}{\diagbox{Model}{Metric}}}

& \multicolumn{2}{c}{\textbf{Distrib. Match}} 
& \multicolumn{4}{c}{\textbf{Alignment}} 
& \multicolumn{4}{c}{\textbf{Aesthetics}} \\
\cmidrule(lr){3-4} \cmidrule(lr){5-8} \cmidrule(lr){9-12}
\multicolumn{2}{l}{} & FAD $\downarrow$ & KL $\downarrow$ 
& CLAP $\uparrow$ &CLaMP3 $\uparrow$ & PER $\downarrow$ & SECS $\uparrow$  
& CE$\uparrow$ & CU$\uparrow$ & PC$\uparrow$ & PQ $\uparrow$  \\
\midrule

\multicolumn{2}{l}{Ground Truth} &- &- &0.18 &0.105  &21.39 &76.42  &7.08 &7.04 &6.41 &7.30 \\
\midrule
\multicolumn{2}{l}{Parler-tts*}  &4.13 &1.00 &0.19 &0.074 &58.61 &64.37 &5.96 &6.62 &5.49 &6.82\\
\multicolumn{2}{l}{Multi-Stage} 	&2.18 &0.78	&0.29 &0.085  &\textbf{38.80} &\textbf{74.04} &6.39	&6.27 &5.90	&6.69\\
\midrule

\midrule

\multirow{2}{*}{\rotatebox{90}{Mixed}}

&Mixed &\underline{1.74} &0.71  &\textbf{0.35} &\underline{0.093}   &51.84  &73.69  &6.50	 &6.66	&\underline{6.14}	&7.03 \\

&\textbf{Mixed pro (ours)}  &\textbf{1.71} &\textbf{0.69} &\textbf{0.35} &\textbf{0.094} &40.58 &\underline{73.78}   &\textbf{6.77}	& \textbf{6.86}	&\textbf{6.18}	&\textbf{7.19}\\
\midrule
\multirow{5}{*}{\rotatebox{90}{Dual-track}}
&Parallel (standard)  &2.45 &0.75 &0.33 &0.087 &48.40 &72.27  &6.21 &6.40 &5.68 &6.88\\
&Parallel (V-A) &2.54 &0.73 &0.33 &0.088 &46.30 &72.43 &6.21 &6.49 &5.72 &6.92\\
&Parallel (A-V)  &2.31 &0.72 &0.34 &0.089 &47.00 &72.50  &6.26  &6.47 &5.80 &6.92\\
 \cmidrule{2-12}
&Interleaving (V-A)  &1.96 &0.71 &0.34 &0.092 &41.82&73.12 &6.52 &6.52 &5.97 &7.03\\
&\textbf{Interleaving (A-V) (ours)}   &1.87 &\textbf{0.69} &\textbf{0.35} &\underline{0.093}  &\underline{39.46} &73.16 &\underline{6.67}	&\underline{6.72}	&6.11	&\underline{7.12}
\\

\bottomrule
\end{tabular}
}
\vspace{-10pt}
\end{center}
\end{table*}

\noindent\textbf{Baselines.} 
To the best of our knowledge, no open-source text-to-song model is currently available.
We therefore conduct a controlled comparison between single-stage and multi-stage approaches. The multi-stage baseline consists of two transformer models with the same architecture and training data as SongGen. Specifically, in Stage 1, the first model generates the vocal track from the lyrics, description, and a 3-second reference voice. In Stage 2, the second model generates the accompaniment, conditioned on the same textual inputs and the generated vocal (prepended into the decoder). The final song is then mixed from the two tracks. 
Additionally, we fine-tune Parler-tts~\cite{lyth2024parler}, a text-to-speech model that generates speech from both transcript and description texts, using our own training data.
We also compare our model with Suno, a commercial product, using human evaluations.

\noindent\textbf{Evaluation dataset and metrics.}
For the evaluation dataset, we filter the English-voiced song samples from MusicCaps benchmark \cite{agostinelli2023musiclm}, yielding a test set of 326 samples, with the lyrics annotated by our preprocessing pipeline. 

For automatic evaluations, Frechet Audio Distance (FAD) measures the generation fidelity; Kullback-Leibler Divergence (KL) evaluates conceptual similarity with the target audio; CLAP Score and CLaMP3 score~\cite{wu2025clamp3}  measures the alignment between the audio and the text description; Speaker Embedding Cosine Similarity (SECS) assesses the similarity of speaker identity; Phoneme Error Rate (PER) gauges adherence to the provided lyrics. Note that due to limitations in the ASR model, the PER values are higher than the actual error rate, but the relative differences remain meaningful. We also introduce content-based aesthetics metrics~\cite{tjandra2025aes}, covering Content Enjoyment (CE), Content Usefulness (CU), Production Complexity (PC), and Production Quality (PQ).
For each method, we generate the audio five times with different random seeds and report the average metric.

For human evaluations, we employ Mean Opinion Score (MOS) tests, assessing five key aspects: overall quality (OVL.), focusing on musicality and naturalness; relevance to the text description (REL.); vocal quality, with an emphasis on clarity and intelligibility of the singing voice (VQ.); harmony between vocals and accompaniment (HAM.); and similarity to the original singer (SS.).
The appendix \ref{sec: eval_detail} shows details of the evaluations.

\subsection{Results of Text-to-song Generation}

Tables~\ref{table:t2s_auto} and~\ref{table:t2s_human} report the automatic and human evaluation results, respectively, for our mixed and dual-track models alongside several baselines.
In both tables, the first 3-second vocal clip from the ground-truth is used as the reference voice for all our models and the multi-stage baseline.

\begin{table}[t]
\caption{Humain evaluation of Text-to-Song generation. The overall first and second results are marked with \textbf{bold} and \underline{underline}, respectively. The top results in both of our generation modes are highlighted in \textcolor{yellow}{yellow}.} 

\label{table:t2s_human}
\begin{center}
\resizebox{1\columnwidth}{!}{
\begin{tabular}{@{}p{0.35cm}@{}p{2cm}@{}ccccc@{}}
\toprule
\multicolumn{2}{l}{Model} & OVL.$\uparrow$ & REL.$\uparrow$ &VQ.$\uparrow$ & HAM. $\uparrow$  &SS. $\uparrow$ \\
\midrule
\multicolumn{2}{l}{Ground Truth} &\textbf{4.57} \scriptsize$\pm$0.04	&\textbf{4.49} \scriptsize$\pm$0.03 	&\textbf{4.49} \scriptsize$\pm$0.05 &\textbf{4.47} \scriptsize$\pm$0.04 	&\textbf{4.58} \scriptsize$\pm$0.03 \\
\multicolumn{2}{l}{Suno} &\underline{4.28}\scriptsize$\pm$0.04 	&3.31\scriptsize$\pm$ 0.04 	&\underline{4.22}\scriptsize$\pm$ 0.05 	&\underline{4.33}\scriptsize$\pm$ 0.05  &-\\

\midrule
\multicolumn{2}{l}{Parler-tts*}  &2.58\scriptsize$\pm$ 0.06 &2.13\scriptsize$\pm$0.05 &2.28\scriptsize$\pm$0.03 &2.35\scriptsize$\pm$0.04 &- \\
\multicolumn{2}{l}{Multi-Stage} &3.39\scriptsize$\pm$0.03 &3.20\scriptsize$\pm$0.04 &3.98\scriptsize$\pm$0.07 &2.97\scriptsize$\pm$0.04 &3.89\scriptsize$\pm$0.03 \\
\midrule
\midrule

\multirow{2}{*}{\rotatebox{90}{Mixed}}
&Mixed  &3.58\scriptsize$\pm$0.05 	&3.70\scriptsize$\pm$0.02 	&3.55\scriptsize$\pm$ 0.07	&3.39\scriptsize$\pm$ 0.05	&3.92\scriptsize$\pm$0.05 \\
&\textbf{Mixed pro}  &\cellcolor[HTML]{FFFF99}3.96 \scriptsize$\pm$0.04 	&3.86\scriptsize$\pm$ 0.04	&4.07\scriptsize$\pm$0.06 	&\cellcolor[HTML]{FFFF99}4.01\scriptsize$\pm$0.05 	&\cellcolor[HTML]{FFFF99}4.04\scriptsize$\pm$0.05 \\
\midrule
\multirow{5}{*}{\rotatebox{90}{Dual-track}}
&Parallel (std.) &3.19\scriptsize$\pm$0.04 &3.27\scriptsize$\pm$0.06 & 3.36\scriptsize$\pm$0.04 &2.98\scriptsize$\pm$0.05 &3.44\scriptsize$\pm$0.03 \\
&Parallel (V-A) &3.36\scriptsize$\pm$0.03 &3.32\scriptsize$\pm$0.05 &3.48\scriptsize$\pm$0.05 &3.08\scriptsize$\pm$0.06 &3.47\scriptsize$\pm$0.04 \\
&Parallel (A-V) &3.40\scriptsize$\pm$0.03 	&3.33\scriptsize$\pm$0.04 	&3.51\scriptsize$\pm$0.04 	&3.21\scriptsize$\pm$0.05 	&3.51\scriptsize$\pm$0.05 \\
 \cmidrule{2-7}
&Inter. (V-A)   &3.77\scriptsize$\pm$0.03 &3.69\scriptsize$\pm$0.05 &3.98\scriptsize$\pm$0.06 &3.65\scriptsize$\pm$0.04 &3.88\scriptsize$\pm$0.04 \\
&\textbf{Inter. (A-V)}   &3.95\scriptsize$\pm$0.03 	&\cellcolor[HTML]{FFFF99}\underline{3.87}\scriptsize$\pm$0.06 	&\cellcolor[HTML]{FFFF99}4.15\scriptsize$\pm$0.05	&3.82\scriptsize$\pm$ 0.03	&3.93\scriptsize$\pm$0.04
\\

\bottomrule
\end{tabular}
}

\end{center}
\end{table}

\noindent\textbf{Comparison with baselines.}
Although Parler-TTS excels in controllable text-to-speech, fine-tuning it for text-to-song  is ineffective, as shown in the tables. This highlights the greater complexity of generating expressive vocals and musically coherent accompaniment in text-to-song tasks, compared to conventional speech synthesis.

Compared to the multi-stage baseline, our single-stage model outperforms it across both automatic and human evaluations. The only exceptions are the PER and SECS metrics, where the multi-stage model performs slightly better—due to its first-stage training being focused exclusively on clean vocal targets.
Nonetheless, our single-stage approach shows clear strengths on all other metrics, particularly in aesthetics-related aspects: CE (+5.9\%), CU (+9.4\%), PC (+4.7\%), and PQ (+7.5\%), as well as in human evaluations, with Overall Quality (OVL.) increasing by 0.57 and Harmony (HAM.) by 1.04 on a five-point MOS scale. 
By directly modeling the joint distribution $ P(\text{vocal}, \text{accompaniment}) $, rather than optimizing $ P(\text{vocal}) $ and $ P(\text{accompaniment} \mid \text{vocal}) $ separately, the single-stage approach facilitates more effective global coordination and avoids error accumulation across stages. This is particularly beneficial in song generation, where the alignment and coherence between vocals and accompaniment are essential. We also showcase generation samples from both the multi-stage and single-stage models on our demo page~\footnote{\url{https://liuzh-19.github.io/SongGen/}}, where the multi-stage outputs are often perceptibly off-beat. In addition to better quality, the single-stage model is also more efficient. On an A800 GPU, our mixed-pro model averages 18.04 seconds to generate a 30-second sample, compared to 42.85 seconds for the multi-stage baseline.

\begin{figure}[tb!]
	\centering
	\includegraphics[width=1\columnwidth]{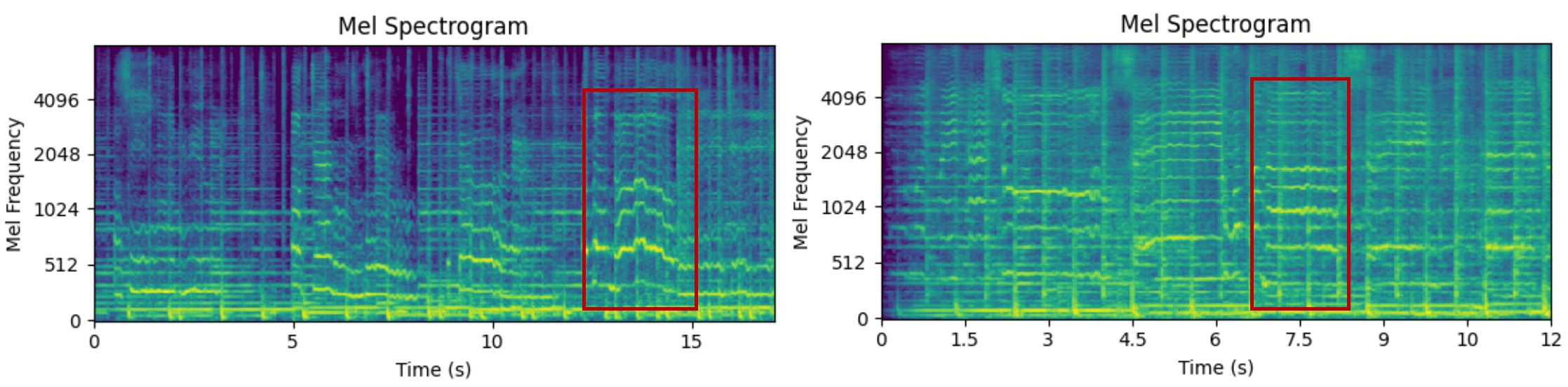}
    \vspace{-15pt}
	\caption{Mel-spectrogram visualization of our generated song featuring various singing techniques. }
	\label{fig:vibrato}
    \vspace{-10pt}
\end{figure}

While SongGen shows some gaps when compared to Ground Truth and Suno, it is worth noting that we use only 2k hours of labeled data, sourced from publicly available datasets. Despite the limited data, SongGen achieves competitive performance. Figure \ref{fig:vibrato} shows a mel-spectrogram of our generated songs, demonstrating that SongGen produces songs with various singing techniques like vibrato. Compared to Suno, a commercial product, SongGen outperforms in terms of text relevance and vocal control. Suno struggles to adhere to the highly detailed textual descriptions in MusicCaps (as shown by the REL. metric) and lacks voice cloning support, giving our model a distinct advantage in these aspects.

\begin{figure}[tb!]
	\centering
	\includegraphics[width=\columnwidth]{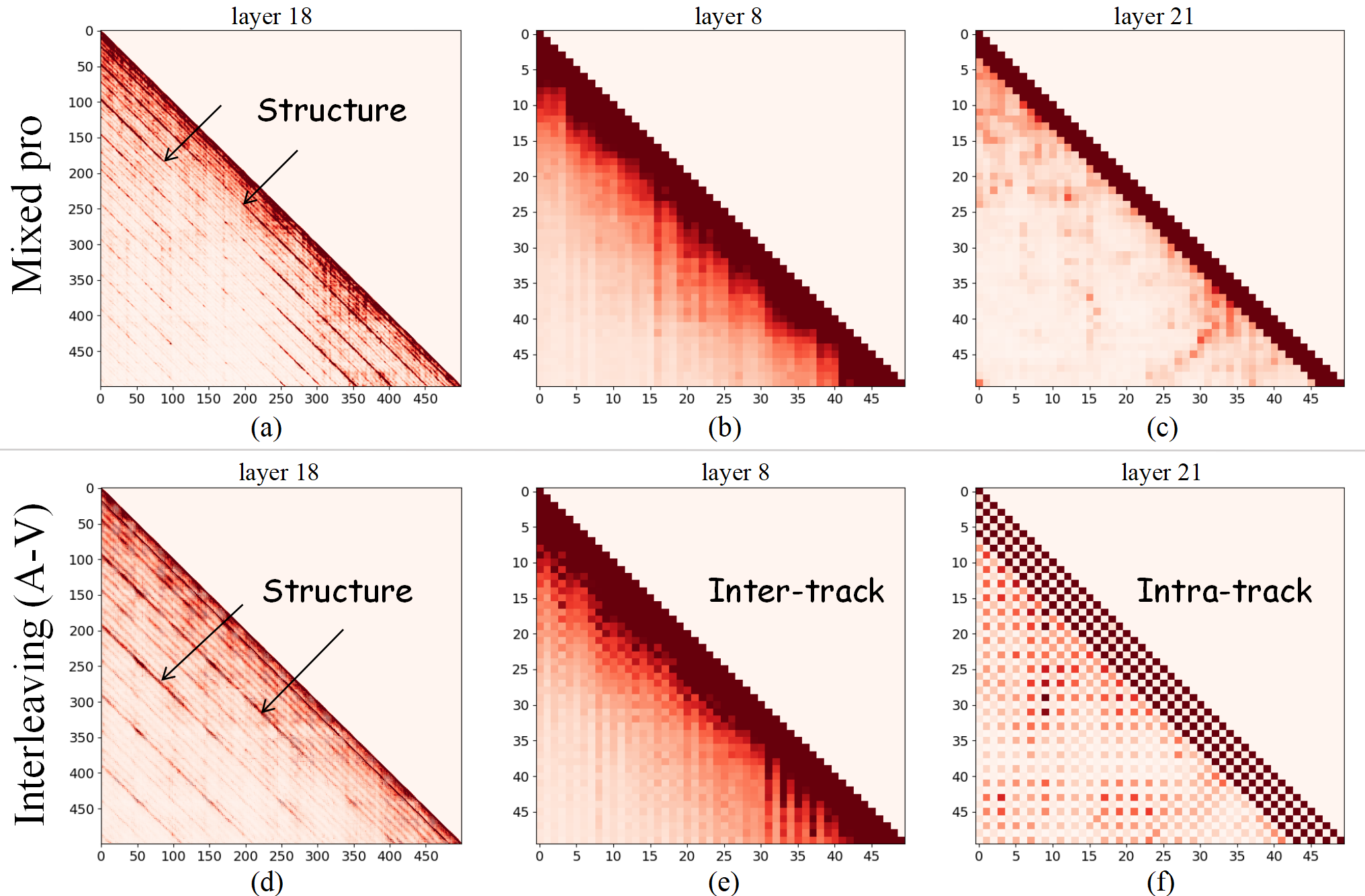}
    \vspace{-15pt}
	\caption{Visualization of decoder attention.}
    \vspace{-10pt}
	\label{fig:attan}
\end{figure}

\textbf{Mixed Mode and Dual-Track Mode.}
We further analyze the performance of the mixed mode and dual-track mode of our framework. 
In mixed mode generation, the "Mixed Pro" approach outperforms the basic "Mixed" model across all metrics, particularly in vocal quality (as indicated by the PER and VQ.).
It indicates that by incorporating an auxiliary vocal token prediction target, the learning biases in mixed mode are effectively mitigated.

 In dual-track mode, the ``Interleaving (A-V)” pattern obtains the best performance.
 Although the parallel pattern is more computationally efficient, its performance lags behind the interleaving pattern. This is likely because, in parallel mode, each hidden state mixes vocals and accompaniment, making separation difficult with only two linear heads. 
 Interestingly, regardless of the pattern (parallel or interleaving), placing the accompaniment before the vocals leads to better results than the reverse order.

Compared to ``Mixed pro",``Interleaving (A-V)" shows competitive performance, with only slightly worse result in FAD. Further comparison reveals that ``Interleaving (A-V)" achieves better vocal quality (VQ.), but its harmony (HAM.) is slightly inferior to that of the ``Mixed pro". This highlights the distinct advantages and challenges of each generation mode. 
We further visualize the attention scores in the decoder to explore the internal mechanisms of the transformer in both modes. Figures \ref{fig:attan} (a),(b), and (c) show self-attention over 500 steps in layer 18 for the ``mixed pro", and over 50 steps in layers 8 and 21. Figures \ref{fig:attan} (a),(b), and (c) present the same for the ``interleaving (A-V)" pattern. From (a) and (d), we observe evenly spaced parallel lines along the diagonal. 
Since songs typically have repetitive structures, this attention pattern suggests that our model has effectively learned the underlying structure of music. Interestingly, in (f), the attention follows a checkerboard pattern, where attention scores for odd steps are strong with other odd steps and similarly, for even steps. This indicates that in the ``interleaving (A-V)" mode, higher layers focus more on learning intra-track relationships, while lower layers (shown in (c)) capture inter-track interactions.

\noindent\textbf{Without a reference voice.}
We explore the song generation capability of SongGen without a reference voice. Table \ref{table:wo_voice} shows that performance declines slightly. However, the listening test results demonstrate that the model continues to produce enjoyable songs with coherent vocals.

\begin{table}[tb!]
\caption{Text-to-Song results without voice input.}

\label{table:wo_voice}
\begin{center}
\resizebox{\columnwidth}{!}{
\begin{tabular}{lccccc}
\toprule
Model & FAD $\downarrow$    & OVL.$\uparrow$ & REL.$\uparrow$ &VQ.$\uparrow$ & HAM. $\uparrow$   \\
\midrule
Mixed  pro &1.96 &3.72\scriptsize$\pm$0.03 	&3.48\scriptsize$\pm$0.04 	&3.88\scriptsize$\pm$0.06 	&3.87\scriptsize$\pm$0.05 \\
Inter.(A-V) &2.21 &3.70\scriptsize$\pm$0.04 &3.47\scriptsize$\pm$0.06 &3.91\scriptsize$\pm$0.04 &3.83\scriptsize$\pm$0.04 \\

\bottomrule
\end{tabular}
}
\vspace{-10pt}
\end{center}
\end{table}

\begin{table}[t]
\caption{Ablation results on training scheme. HQFT is short for High-Quality Finetuning and CL stands for curriculum learning.}
\label{table:loss}
\begin{center}
\resizebox{0.95\columnwidth}{!}{
\begin{tabular}{lccccc}
\toprule
Model & FAD $\downarrow$ & KL $\downarrow$ & CLAP $\uparrow$   & PER $\downarrow$ & SECS $\uparrow$  \\
\midrule
w/o HQFT &2.01 &0.72 &0.32 &43.68 &72.83 \\
w/o CL &2.35 &0.73 &0.33 &55.71  &72.81 \\
\midrule
\textbf{ours}  &\textbf{1.71} &\textbf{0.69} &\textbf{0.35} & \textbf{40.58} &\textbf{73.78}\\ 
\bottomrule
\end{tabular}
}
\vspace{-15pt}
\end{center}
\end{table}

\begin{table}[t]
\caption{Ablation results on different lyric integration methods.}
\vspace{-10pt}
\label{table:lyrics}
\begin{center}
\resizebox{\columnwidth}{!}{
\begin{tabular}{ccccccccc}
\toprule
Tokenizer & \makecell[c]{w/ Lyrics \\Encoder} & \makecell[c]{prepend \\/ cross} & FAD $\downarrow$ & PER $\downarrow$ & SECS $\uparrow$ \\
  
\midrule
VoiceBPE & \xmark &prepend   &3.41   &62.38 &69.09\\
VoiceBPE & \cmark &prepend &3.56   &56.21 &70.70\\ 
VoiceBPE &\xmark &cross &1.95  &61.81 &72.59\\
T5  &\cmark  &cross &1.88  &55.27 &73.67\\
\midrule
VoiceBPE &\cmark &cross &\textbf{1.73}  &\textbf{43.34} &\textbf{73.59}\\
\bottomrule
\end{tabular}
}
\vspace{-10pt}
\end{center}
\end{table}

\subsection{Ablation Studies}
In this section, we conduct extensive ablation studies. Since both mode are based on a unified framework, we present results from the mixed mode setting due to space limitations.

\noindent\textbf{Effect of training strategy.}
In Table \ref{table:loss}, we evaluate the effectiveness of our High-Quality Finetuning (HQFT) and curriculum learning (CL) strategy for codebook loss weights. HQFT improves all metrics, confirming the effectiveness of our quality filtering criteria. 
Compared to the ``w/o CL" variant, where each codebook's loss weight is fixed and equal, our CL strategy improves performance. This demonstrates that prioritizing the most important tasks first and then progressively refining the details is effective.

\noindent\textbf{Effect of Lyrics Module Design.}

We further investigate the impact of different lyric integration methods, including the choice of tokenizer (VoiceBPE vs. T5), the use of a lyrics encoder, and the integration approach (pre-pending vs. cross-attention). Figure \ref{table:lyrics} shows the results after Step 1 training for each variant. Our design (VoiceBPE, w/ lyrics encoder, cross-attention) achieves the best results across all metrics, validating the effectiveness. Unlike most TTS works, which prepend transcripts before audio tokens, we find that the cross-attention approach is more effective and stable. This may be because cross-attention allows the decoder to focus solely on generating the audio modality. Additionally, phoneme-like tokenizer (VoiceBPE) is more suitable than the word-level tokenizer (T5) for song lyric tokenization. Under this mechanism, the lyrics encoder can capture the relationships between lyric tokens, learning pronunciation patterns from different token combinations, and thus alleviating the burden of modality alignment on the decoder.

\section{Limitations and Future Work}
We acknowledge the limitations of our proposed SongGen model. Due to the scarcity of open-source song data, the current model can only generate songs up to 30 seconds in length, which is insufficient for producing songs with complete structures. Additionally, the current audio codec, X-Codec, operates at a sampling rate of 16kHz. To improve fidelity, our future work will involve training a renderer to upsample the audio for higher quality output.

\section{Conclusion}
In this paper, we introduced SongGen, a fully open-source, single-stage auto-regressive transformer for text-to-song generation. Operating within a unified framework, we devised a variety of token patterns. These patterns endow SongGen with the ability to support two distinct generation modes: the mixed mode and the dual-track mode. Experimental outcomes convincingly demonstrate the efficacy of our token pattern design. Moreover, they showcase the strong song generation capabilities of SongGen in both the mixed mode and the dual-track mode.

\section*{Acknowledgments}
This work was supported by National Key R\&D Program of China 2022ZD0161600, Shanghai Artificial lntelligence Laboratory, the Centre for Perceptual and Interactive Intelligence (CPII) Ltd under the Innovation and Technology Commission (ITC)’s InnoHK. Dahua Lin is a PI of CPII under the InnoHK.

\clearpage

\section*{Impact Statement}
The proposed work, SongGen, a controllable text-to-song generation model, has the potential to impact various aspects of society. On the positive side, SongGen enables both content creators and novices to effortlessly express their creativity with a low entry barrier, while also streamlining the workflow for experienced music producers.

However, since SongGen autonomously generates songs and supports voice cloning, there are risks of copyright infringement, intellectual property misuse, and the creation of deepfake audio. Proper constraints are needed to ensure the model is not misused in illegal or unethical ways.

In conclusion, while SongGen presents exciting possibilities for the music industry and creative expression, its development should be accompanied by careful consideration of its ethical and societal implications.


\bibliography{example_paper}
\bibliographystyle{icml2025}

\newpage
\appendix
\onecolumn

\section{Implementation and Training Details}
In SongGen, the lyrics encoder is a 6-layer transformer with a hidden size of 1024. The transformer decoder, consisting of 24 layers with 1024 hidden size, includes both causal self-attention and cross-attention blocks in each layer. In ``Mixed Pro" Mode, the vocal loss weight $\lambda$ is set to 0.1. The model is trained for approximately 400K steps using 16 Nvidia A100 (80GB) GPUs, with a batch size of 16 per GPU. For optimization, we employ the AdamW optimizer~\cite{loshchilov2018decoupled} with $\beta_1 = 0.9$, $\beta_2 = 0.99$, and a weight decay of $10^{-4}$. During training step 1, the learning rate is set to $10^{-4}$, while for the subsequent fine-tuning steps, the learning rate is reduced to $5 \times10^{-5}$. A cosine learning rate schedule is applied for all traning steps. 
The resource allocation and training strategy for the multi-stage baseline are consistent with those used for SongGen mixed training Step 1. Since voice-free support is not directly related to the core comparison between the single-stage and multi-stage designs, we omit this part of the multi-stage model. Specifically, the two models in the multi-stage pipeline are trained separately for approximately 200K steps each. We observe that the loss begins to plateau around 60K steps for both models. 
To facilitate reproducibility, we make our training configurations publicly available.

\section{Details in Evaluations}
For evaluation, we select 326 samples from the MusicCaps \cite{agostinelli2023musiclm} benchmark, with no overlap with the training set. MusicCaps test set contains 2.8K samples, with captions written by expert musicians. However, many of the samples are instrumental music, sound effects, or speech. We filter the English-voiced song samples from MusicCaps test set using our automated data preprocessing pipeline, resulting a final set of 326 samples. Note that the evaluation set was selected impartially, with no intention to influence fairness.

\label{sec: eval_detail}
For objective metrics, all samples are normalized at-14dB LUFS for fairness.
Frechet Audio Distance (FAD)~\cite{kilgour2019fad} evaluates the fidelity of generated songs by calculating the distribution distance between features of the target and generated audio, extracted from the VGGish~\cite{hershey2017vggish} model.
Kullback-Leibler Divergence (KL) measures the similarity between the generated and target audio with the label calculated by the audio tagging model. A lower KL suggests that the generated music shares similar concepts with the reference. 
FAD and KL are calculated using \texttt{audioldm\_eval}~\footnote{\url{https://github.com/haoheliu/audioldm_eval}}.
CLAP Score evaluates the alignment between generated audio and the given text prompt using the official CLAP model~\cite{laionclap2023}, implemented via the \texttt{stable-audio-metrics}~\footnote{\url{https://github.com/Stability-AI/stable-audio-metrics}}. However, we observe that the CLAP Score for MusicCaps test set is unexpectedly low. We supplement our evaluation with the CLaMP3 Score~\cite{wu2025clamp3}, a more recent model that provides a more robust measure of text–song alignment.
Phoneme Error Rate (PER) assesses the adherence of the generated audio to the provided lyrics by transcribing the audio using Distill Whisper\cite{gandhi2023distilwhisper} and computing the phoneme error rate against the reference lyrics. However, PER is not an ideal measure of vocal quality, as current ASR models struggle with sung vocals.
Speaker Embedding Cosine Similarity (SECS) assesses the similarity of speaker identity using the Resemblyzer\footnote{\url{https://github.com/resemble-ai/Resemblyzer}} speaker encoder to compute the SECS between reference 3-second vocal clips and generated audio.
We further introduce recently proposed content-based aesthetics metrics~\cite{tjandra2025aes}, covering Content Enjoyment (CE), Content Usefulness (CU), Production Complexity (PC), and Production Quality (PQ).

For the subjective evaluations, we randomly select 36 audio samples generated by our models, and each sample is evaluated by 20 listeners. We conduct the commonly used MOS (Mean Opinion Score) tests across five aspects. The rating scale ranges from 1 to 5, with higher scores indicating better performance.
For the Overall Quality (OVL.) evaluation, we instruct the raters to focus on musicality and naturalness, while ignoring style differences.
For the Relevance to Text Description (REL.) evaluation, we ask the raters to score based on the proportion of key points from the text description that are reflected in the generated song.
For the Vocal Quality (VQ.) evaluation, we emphasize the importance of clarity, lyric accuracy, and the naturalness and coherence of the vocals in the ratings.
For Harmony (HAM.), we ask the raters to pay particular attention to the temporal correspondence between the accompaniment and the vocals.
For Speaker Similarity, we ask the raters to focus on the similarity of the speaker's identity (timbre) to the reference, ignoring differences in content.
A small subset of the samples used in the test is available on our project page \url{https://liuzh-19.github.io/SongGen/}.

\section{The Impact of Different Audio Codecs.}

\begin{figure*}[tb!]
	\centering
	\begin{minipage}{0.49\textwidth}
		\centering
		\begin{table}[H]
			\caption{Ablation results of different neural audio codecs.}
			\label{table:codec}
			\begin{center}
				\resizebox{1\columnwidth}{!}{
					\begin{tabular}{lccccc}
						\toprule
						Model & FAD $\downarrow$ & KL $\downarrow$ & CLAP $\uparrow$ & PER $\downarrow$ & SECS $\uparrow$ \\
						\midrule
						Encodec &10.84 &0.99 &0.19 &60.67 &71.36\\
						DAC &4.36 &0.86 &0.24 &68.64 &71.66\\
						\midrule
						X-Codec (ours) &\textbf{1.73} &\textbf{0.70} &\textbf{0.33} &\textbf{43.34} &\textbf{73.59} \\
						\bottomrule
					\end{tabular}
				}
			\end{center}
		\end{table}
	\end{minipage}
	\hfill
	\begin{minipage}{0.48\textwidth}
		\centering
		\includegraphics[width=0.8\textwidth]{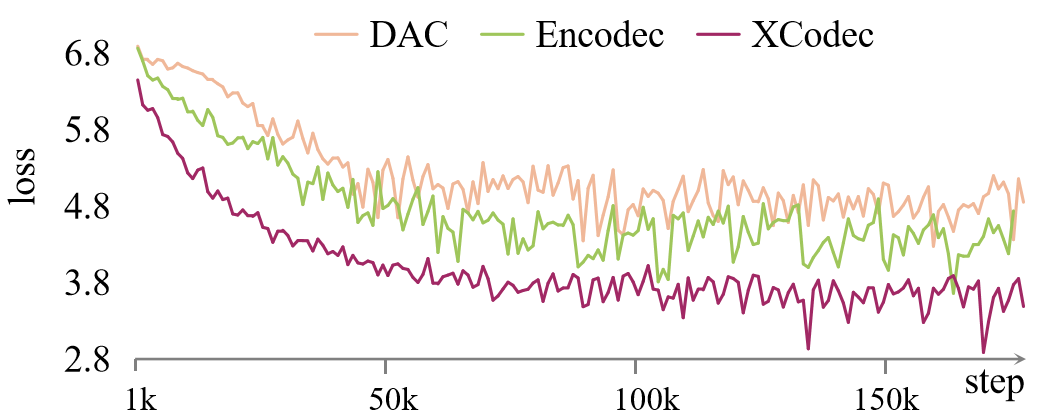}
            \vspace{-6pt}
		\caption{Training loss curves of different audio codecs}
		\label{fig:codec}
	\end{minipage}
\end{figure*}

We compare the performance of three different codecs: XCodec, Encodec (24kHz)~\cite{defossez2022encodec}, and DAC (44.1kHz)~\cite{kumar2024DAC}. 
Table \ref{table:codec} shows the results after training Step 1. X-Codec surpasses both Encodec and DAC on all metrics. Additionally, the loss curves in Figure \ref{fig:codec} demonstrate that X-Codec exhibits more stable training and faster convergence. 
Although Encodec and DAC have been widely adopted in prior audio generation systems across domains such as speech~\cite{lyth2024parler} and instrumental music~\cite{copet2024musicgen}, song generation presents a substantially higher level of semantic complexity.
While Encodec and DAC indeed yield better perceptual quality for audio reconstruction, we observed that in song generation, both codecs resulted in higher rates of invalid outputs, such as failure to follow lyrics, or producing noise and silence. In contrast, X-Codec consistently demonstrated more stable training, faster convergence, and higher success rates in generating coherent vocals. 

We attribute this performance to several factors. First, we adopt the publicly released \texttt{xcodec\_hubert\_general\_audio} checkpoint, trained on a large-scale (200k-hour) private dataset with a distribution similar to AudioSet~\cite{gemmeke2017audioset}. We speculate that its exposure to large amounts of music data during pretraining contributes to its superior performance in our task. Second, as emphasized in the X-Codec paper~\cite{ye2024xcodec}, the incorporation of not only acoustic but also semantic features from self-supervised learning representations might also contribute to the performance. 
Despite X-Codec operating at a relatively low sampling rate of 16 kHz, we selected it as the most suitable option available at the time. To date, high-fidelity, song-specific neural codecs tailored for generative modeling remain an open challenge in the research community.

\end{document}